\documentclass{emulateapj}
\usepackage{epsfig,color}
\usepackage{natbib}

\def\ltsima{$\; \buildrel < \over \sim \;$}
\def\simlt{\lower.5ex\hbox{\ltsima}}
\def\gtsima{$\; \buildrel > \over \sim \;$}
\def\simgt{\lower.5ex\hbox{\gtsima}}

\def\lesssim{\mathrel{\hbox{\rlap{\hbox{\lower4pt\hbox{$\sim$}}}\hbox{$<$}}}}
\def\gtrsim{\mathrel{\hbox{\rlap{\hbox{\lower4pt\hbox{$\sim$}}}\hbox{$>$}}}}

\newcommand\ledd{{L}_{\rm Edd}}

\def\msun{{\,{\rm M}_\odot}}

\newcommand\mbh{{\,{\rm M}_{\rm bh}}}

\def\del#1{{}}

\begin{document}

\title{The observed $M - \sigma$ relations imply that SMBHs grow by cold chaotic accretion.}


\author{Sergei Nayakshin\altaffilmark{1}, Chris Power\altaffilmark{2} \&
  Andrew R. King\altaffilmark{1}}
\affil{ $^1$ Department of Physics \& Astronomy, University of Leicester, 
          Leicester, LE1 7RH, UK\\
	  $^2$ International Centre for Radio Astronomy Research, University 
          of Western Australia, 35 Stirling Highway, Crawley, WA 6009, Australia}





\begin{abstract}
 We argue that current observations of $M - \sigma$ relations for galaxies can
 be used to constrain theories of super-massive black holes (SMBH) feeding. In
 particular, assuming that SMBH mass is limited only by the feedback on the
 gas that feeds it, we show that SMBHs fed via a planar galaxy scale gas flow,
 such as a disc or a bar, should be much more massive than their counterparts
 fed by quasi-spherical inflows. This follows from the relative inefficiency
 of AGN feedback on a flattened inflow. We find that even under the most
 optimistic conditions for SMBH feedback on flattened inflows, the mass at
 which the SMBH expels the gas disc and terminates its own growth is a factor
 of several higher than the one established for quasi-spherical inflows. Any
 beaming of feedback away from the disc and any disc self-shadowing
 strengthens this result further. Contrary to this theoretical expectation,
 recent observations have shown that SMBH in pseudobulge galaxies (which are
 associated with barred galaxies) are typically under- rather than
 over-massive when compared with their classical bulge counterparts at a fixed
 value of $\sigma$. We conclude from this that SMBHs are not fed by large (100
 pc to many kpc) scale gas discs or bars, most likely because such planar
 flows are turned into stars too efficiently to allow any SMBH growth. Based
 on this and other related observational evidence, we argue that most SMBHs
 grow by chaotic accretion of gas clouds with a small and nearly randomly
 distributed direction of angular momentum.
\end{abstract}


\section{Introduction}\label{intro}

It is now well established that super-massive black holes (SMBHs) reside in
the nuclei of many galaxies. The masses of these SMBHs correlate with a range
of properties of their host spheroids, including luminosity
\citep[e.g.][]{magorrian98} and consequently mass
\citep[e.g.][]{MarconiHunt03,Haering04}; the concentration of stellar bulge
light \citep[e.g.][]{GrahamEtal01}; the deficit of stellar bulge light in
ellipticals \citep[e.g.][]{Graham04,FerrareseEtal06a,Kormendy&Bender09}; and
the gravitational binding energy of the bulge
\citep[e.g.][]{Feoli05,deFrancesco06,Feoli07,Aller&Richstone07}.

The correlation between SMBH mass ($M_{\rm bh}$) and the velocity
  dispersion ($\sigma$) of its host spheroid, referred to below as a ``$M_{\rm
    bh}$--$\sigma$ relation'', has been studied by many authors. A power-law
  fit to the data, $M_{\rm bh} \propto \sigma^p$, yielded values of $p$ in the
  range of $p\sim 4-5$ 
\citep[cf.][]{Ferrarese00,Gebhardt00,Tremaine02,GultekinEtal09}, and is now
widely interpreted as the imprint of the self-regulated growth of the
SMBH. A more recent extended sample of galaxies shows that the power law
  index $p$ may be as high as $p\approx 6$ \citep{GrahamEtal11}.

Furthermore, recent work has highlighted that the $M_{\rm bh}$--$\sigma$
relation one measures will depend on the nature of the galaxy sample.  For
  example, \cite{McConnellEtal11a} presents two extremely massive SMBH, $\mbh
  \sim 10^{10}\msun$, in the centres of two giant elliptical
  galaxies. Combined with other previous SMBH mass measurements, these authors
  obtain $p\approx 4.5$ for elliptical and S0 galaxies. A similar but offset
  to lower SMBH masses result is found for spiral galaxies in the
  sample. Alternatively, when data for all galaxies are combined in one
  $M_{\rm bh}$--$\sigma$ relation, $p\approx 5$ is found.

 Another important conclusion is gradually emerging from the observations:
  SMBHs in galaxies that show flattened morphology, such as discs,
  pseudobulges or bars appear to be less massive at same $\sigma$ than their
  counterparts in elliptical galaxies. In particular, \cite{Hu08} showed that
pseudobulge galaxies tend to have undermassive SMBHs, and \cite{Graham08}
showed that barred galaxies tend to have SMBHs that are a factor of $\sim 3$
to ten smaller than elliptical galaxies at the same velocity dispersion. While
\cite{KormendyEtal11}, in variance with other authors, find that SMBHs
in pseudobulges and galaxy discs do {\em not} even follow an $M_{\rm
  bh}$--$\sigma$ relation, their results are consistent with that of the
  other authors in that most of their pseudobulge systems lie below the
  $M_{\rm bh}$--$\sigma$ for ellipticals.  \cite{MathurEtal12} firmed
  up these conclusions with a sample of narrow line Seyfert 1 nuclei (all
classifiable as pseudobulge systems). A useful compilation of the recent data,
summarising these points, is given in the right panel of Fig. 1 in
\cite{ShankarEtal12}.

The purpose of our paper is to show that there is a unified theoretical
picture within which this observational trend can be understood, and
that the observed $M_{\rm bh}$--$\sigma$ relations place very interesting
constraints on the still unknown feeding mode of SMBHs.

A well known difficulty in fuelling active galactic nuclei (AGN) is
that the typical angular momentum of gas in the galactic bulge is very
large compared with that of the last stable orbit around a black hole
\citep[e.g.,][]{Krolik99,Combes01,Jogee04}. One expects that due to
circularisation shocks, gas will end up in a planar feature, such as a
disc or a ring \citep[cf. simulation S30 shown in Figures 1 and the
left panel of Figure 13 in][]{HobbsEtal11}.  It has been argued that
large scale angular momentum transfer occurs rapidly enough through
the action of stellar and gaseous bars \citep[e.e.,][]{Shlosman90} or
via spiral density waves \citep{Thompson05} to overcome this
difficulty. These galaxy-wide gas discs (with radial scales of many
kpc) could then extend all the way to the inner galaxy and even feed
the SMBH. This ``planar mode'' of feeding is schematically illustrated
in the left panel of Figure 1.

However, gas discs tend to become unstable to gravitational
fragmentation and resulting star formation (SF) beyond a self-gravity
radius of $\sim 0.1$ pc
\citep[e.g.,][]{Paczynski78,Kolykhalov80,Shlosman89,Collin99}. In the
simplest model this would result in complete consumption of gas in the
disc by star formation \citep{NayakshinEtal07} and so there would be
no way to fuel the SMBH. However, stellar feedback could heat up the
large scale disc to self-regulate the star formation rate in the disc
to a level such that a sufficient amount of gas can reach the SMBH,
but there is no clear consensus on this point
\citep[e.g.,][]{Goodman03,Thompson05}.

An alternative solution to the difficulty of fuelling AGN is provided by chaotic feeding
\citep{KingPringle07} or ballistic accretion of cold streams when the gas is
cold enough to form dense clouds or filaments \citep{NK07,HobbsEtal11}. Large
scale cosmological simulations show that cold streams may be important for supplying 
relatively low angular momentum gas to galaxies\citep[e.g.,][]{KeresEtal2009,KimmEtal2011,DuboisEtal2011} , 
while higher resolution simulations that resolve the regions nearer to
SMBH suggest that similar processes may operate on smaller scales
\citep{LevineEtal10}. In particular, the numerical experiments of
\cite{HobbsEtal11} show that turbulence (driven by stellar feedback in the bulge)
in a quasi-spherical distribution of gas leads to formation of convergent
flows that create high density filaments. The latter can travel almost
ballistically through the rest of the bulge, with some filaments arriving in
the inner parsecs of the galaxy. Note that most of the gas still ends up in a
galaxy-scale disc with a large angular momentum; however, the SMBH receives
most of its fuel not from that disc but from the gas filaments arriving at its
vicinity directly.

This ``chaotic cold streams'' feeding mode is schematically shown in the right
panel of Figure 1. In this class of model the orientation of the innermost
sub-pc scale disc makes a random walk because of the arrival of material with a
fluctuating angular momentum. The cancellation of the angular momentum in
shocks transfers gas into the SMBH more rapidly than viscous torques can, and so 
these discs should be more resilient to star formation. Furthermore, because
the orientation of the inner disc is constantly changing, feedback from the AGN 
will be quasi-isotropic.\\

The main point of our paper can be summarised as follows. Large scale gas discs 
are better able to withstand AGN feedback than quasi-spherical inflows
\citep{NayakshinPower10}. This is so even under the most optimistic
assumptions about AGN feedback -- that is, feedback acting directly on the disc 
without being shadowed by the inner disc, and no beaming away from 
the disc/galaxy plane. We show that, under these most favourable of conditions
for which AGN feedback is maximal, there is a critical SMBH mass at which its 
feedback expels the disc that feeds it. We call this a ``disc $\mbh$--$\sigma$ 
relation'' because it arises only for SMBHs accreting from large scale galaxy discs. 
This mass is larger than the canonical $\mbh$--$\sigma$ relation, and any dilution 
of AGN feedback would increase it even further. Theoretically we see that if SMBHs were
fed by planar accretion then we could expect SMBHs in pseudobulge systems to lie 
above the $\mbh$--$\sigma$ relation for classical
bulges. Observationally there is no strong
evidence for this \citep[e.g.][]{Hu08,Graham08,KormendyEtal11}, and so we conclude that most 
SMBHs do not grow via the planar accretion mode but are much more likely to be 
fed by quasi-spherical inflows as in the models of \citet{KingPringle07} and 
\citet{HobbsEtal11}.

 To avoid misunderstanding, we note that galaxy-scale gas discs and
  gas bars are obviously important for galaxy evolution as a whole, e.g., as a
  birthplace for a significant fraction of all stars, and also by effecting a
  radial redistribution of gas and stars within the galaxy, but we see no
  clear observational evidence that these features are able to channel their
  gas all the way to the SMBHs.

\section{The gas weight argument for a spherical AGN
  feedback}\label{sec:spherical}

We briefly review existing ideas of AGN feedback acting on
quasi-spherical distributions of gas before moving on to the problem of AGN
feedback acting on a gas disc. \cite{SilkRees98} assumed feedback in the form of an
energy-conserving outflow and obtained a scaling $M_{\rm
  bh}\propto\sigma^5$. Subsequently, \citet{Fabian99}, \cite{King03,King05} and
  \citet{MurrayEtal2005} solved the momentum equations for
momentum-conserving AGN feedback acting on a
spherical shell of gas in an isothermal bulge potential (see below) and showed
that it leads to a $M_{\rm bh}$--$\sigma$ relation 
\begin{equation}
M_{\rm bh} = {f_g\kappa\over \pi G^2}\sigma^4,
\label{msig}
\end{equation}
where we follow King's analysis; here $\kappa$ is the opacity, which
is assumed to be dominated by the electron scattering, and $f_g$ is
the baryon fraction, which is assumed to be equal to the initial
cosmological value of $f_g \simeq 0.16$ \citep[cf.][]{SpergelEtal07}.
 We note that all of the aforementioned theoretical approaches used a
singular isothermal sphere potential \citep[e.g., \S 4.3.3b
in][]{BT08} for simplicity. For such a potential, the one dimensional
velocity dispersion is a constant independent of radius, $\sigma =
(GM_{\rm total}/2R)^{1/2}$, where $M_{\rm total}(R)$ is the total
enclosed mass including dark matter inside radius $R$ (the distance
from the centre of the galaxy). The enclosed gas mass, $M(R) = f_g M_{\rm
  total}(R) = 2 f_g \sigma^2 R/G$ is proportional to $R$. The gas
density at radius $R$ for such a potential is $\rho_g(R) = f_g
\sigma^2/(2\pi G R^2)$ for reference.

\cite{King03,King05} assumed that the SMBH
luminosity is limited by the Eddington value $L_{\rm Edd}$, and the momentum
outflow rate produced by radiation pressure \citep{KP03} is of order
\begin{equation}
\dot\Pi_{\rm SMBH} \approx {L_{\rm Edd}\over c} = \frac{ 4 \pi G M_{\rm bh}}{\kappa}\;.
\label{pismbh}
\end{equation}
We can recover the result of \cite{King03,King05} using a simpler
``weight argument'', which requires that the momentum output produced
by the SMBH ($\dot\Pi_{\rm SMBH} $) should just balance the weight of
the gas in the bulge, $W(R) = GM(R)[M_{\rm total}(R)]/ R^2$, which
turns out to be independent of radius:  
\begin{equation}
W = { GM(R) M_{\rm total}(R) \over R^2} = {4f_g\sigma^4\over G}\;.
\label{w}
\end{equation}
To order of magnitude, equation \ref{w} holds for any potential if estimated
at a radius which encloses most of the gas potentially available for SMBH
fueling, e.g., such as the virial radius for the \cite{NFW}
potential. Balancing the outward force of the outflow (equation \ref{pismbh})
with the weight of the gas in the bulge (equation \ref{w}) then leads
naturally to equation~\ref{msig}.

The momentum feedback model is attractive in its physical simplicity. The
model is underpinned by observations of fast $v \sim 0.1-0.3 c$ outflows from
bright AGN and quasars \citep[e.g.,][] {PoundsEtal03b,PoundsEtal03a}. Studying
the blue shifted ionised absorption features in the 7 to 10 keV energy range
in a large sample of AGN with {\em XMM-Newton}, \cite{TombesiEtal10} concluded
that such fast AGN outflows appear in $\sim 40$ \% of the sample. As these
authors emphasise, their uniform sample overcomes ``publication bias'' and
shows that fast AGN outflows are widespread. It also requires the outflows to
be wide-angled (rather than jet-like), which is beneficial in spreading the
influence of AGN feedback as broadly over the galaxy bulge as possible.

Equation \ref{msig} contains no free parameters, but nevertheless it is very
close to the observed $M_{\rm bh}$--$\sigma$ relation. Similar reasoning helps
to explain both the observed $M_{\rm bh}$--$\sigma$ relation for Nuclear star
Clusters \citep[see][]{McLaughlinEtal06} and the observation that NCs are
preferentially found in low mass (low $\sigma$) galaxies
\citep[][]{NayakshinEtal09b}. The model has been used recently by
\citet{ZubovasEtal11a} to explain the two $\sim$10 kpc scale symmetric lobes
apparently filled with cosmic rays detected by {\em FERMI-LAT} telescope in
the Milky Way \citep{SuEtal10}. To explain the particular geometry of the
bubbles, the only adjustment to the basic \cite{King05} model required by
\cite{ZubovasEtal11a} has been an addition of a dense disc of molecular gas,
known as the Central Molecular Zone, found in the central $\sim$ 200 pc of our
Galaxy \citep{MorrisSerabyn96}.

 While it is clear that the exact $\mbh-\sigma$ relation derived
  analytically is affected by the choice of the potential, we note that the
  main conclusion of our paper -- the fact that it is harder to expel a disc
  feeding the SMBH rather than a quasi-spherical shell -- remains unchanged as
  it is based on a simple geometrical argument (see below). Furthermore,
  gravitational potentials of elliptical and lenticular galaxies appear to be
  relatively well approximated by a constant $\sigma$ value inside their
  effective radii. \cite{CappellariEtal06} find that the 1D velocity
  dispersion of stars within the galaxies in their sample depends on radius
  $R$ (within the galaxy) as a very weak power-law: $\sigma(R)\propto
  R^{-0.066}$. This implies that $\sigma$ drops on average by the factor of
  $\approx 1.16$ when $R$ changes by an order of magnitude, and $\sigma^4$
  varies by the factor of $\approx 1.8$.

Indeed, \cite{McQMcL12a} consider SMBH momentum feedback for potentials other
than isothermal, e.g., the NFW potential \citep{NFW}, \cite{Hernquist90} and
\cite{DWMD05} potentials.  In all cases momentum driven outflow produce
relations of the form $\mbh \propto V_{\rm m}^4$, where $V_{\rm m}$ is the
maximum circular velocity in the potential. The normalisation of the relation
is slightly smaller than for the simpler isothermal potential we use here,
which implies that escape from the isothermal potential is slightly more
demanding than for all the others. To quote a numerical factor here,
\cite{McQMcL12a} find a change in the critical SMBH mass (only) by a factor of
order unity only for the NFW potential, as an example, if one defines $\sigma$
at the peak of the rotation curve of the potential as $\sigma = V_{\rm
  m}/\sqrt{2}$. 

No analytical treatment for the complex problem at hand can be expected to be
accurate within better than a factor of $\sim 2$, and therefore we feel that
the singular isothermal potential is an appropriate approximation to use here.

\section{The theoretical disc ${\mathbf M_{\rm bh}-\sigma}$ relation}\label{sec:msigma}


As shown by \cite{NayakshinPower10}, the above argument does not quite
work for a non-spherical geometry. To see this, consider the simplest
example -- an axially symmetric rotating galaxy. The gas settles in
the plane of symmetry of the galaxy and forms a rotationally supported
disc. Let $H$ be the local (i.e., at a given $R$) vertical disc
scale-height. Let us consider a disc at radius $R$. The {\em maximum}
momentum flux from the SMBH outflow striking the disc is
\begin{equation}
\dot\Pi_{\rm disc} = \frac{H}{R} \;\frac{\ledd}{c}\;.
\label{pdisc}
\end{equation}
This is an absolute maximum given by the total momentum outflow rate (equation
2) times the fraction of the solid angle the disc subtends as seen from the
SMBH location. In reality some of the momentum outflow from the AGN could be
intercepted (shadowed) by the disc at smaller radii (cf. the left panel of
Figure 1), and also the outflow may be beamed along the direction
perpendicular to the disc. These individual effects actually strengthen 
the conclusion we draw because they act to lessen the effect of feedback on the
disc and therefore should increase the mass of the SMBH.

We first assume that the disc mass is dominated by gas, and we
  correct for the presence of stars {\bf within the disc} later on. The effective weight
of the gas disc is given by $W_{\rm eff} \sim GM_{\rm total}(R)M_{\rm
  disc} /R^2$, where $M_{\rm disc}(R) \equiv \pi R^2 \Sigma(R)$ is the
disc mass, and $\Sigma(R) = 2 H(R) \rho_c(R)$ is the gas disc surface
density, with $\rho_c(R)$ being the mid-plane disc density. We note
here that a disc in circular rotation actually has a zero weight
because gravity is exactly balanced by the centrifugal force. However,
within a factor of a few the effective weight defined above still
applies\footnote{This can be seen by observing that escape velocity
  from distance $R$ to infinity for a circularly rotating disc is
  smaller by a factor of $\sqrt{2}$ than that from a static point
  within the same potential} if we wish to expel the disc to infinity
(where the centrifugal force becomes negligible if the angular
momentum of the material is conserved). Therefore, by requiring that
momentum flux balances weight ($\dot\Pi_{\rm disc} \simlt W_{\rm
  eff}$) we limit the SMBH mass to
\begin{equation}
M_{\rm bh} = {\kappa \over 4 \pi} \; {M_{\rm total}(R) M_{\rm disc} \over RH} = {\kappa
  \sigma^2 \over 2 \pi G}\frac{M_{\rm disc}}{H}\;.
\label{mmax0}
\end{equation}
A casual look at this equation would seem to suggest that, because $H$ and $M_{\rm disc}$ may vary with 
radius $R$ and from system to system, one should not expect any robust correlation between $M_{\rm bh}$ 
and $\sigma^4$ to arise.

\begin{figure*}
\centerline{\psfig{file=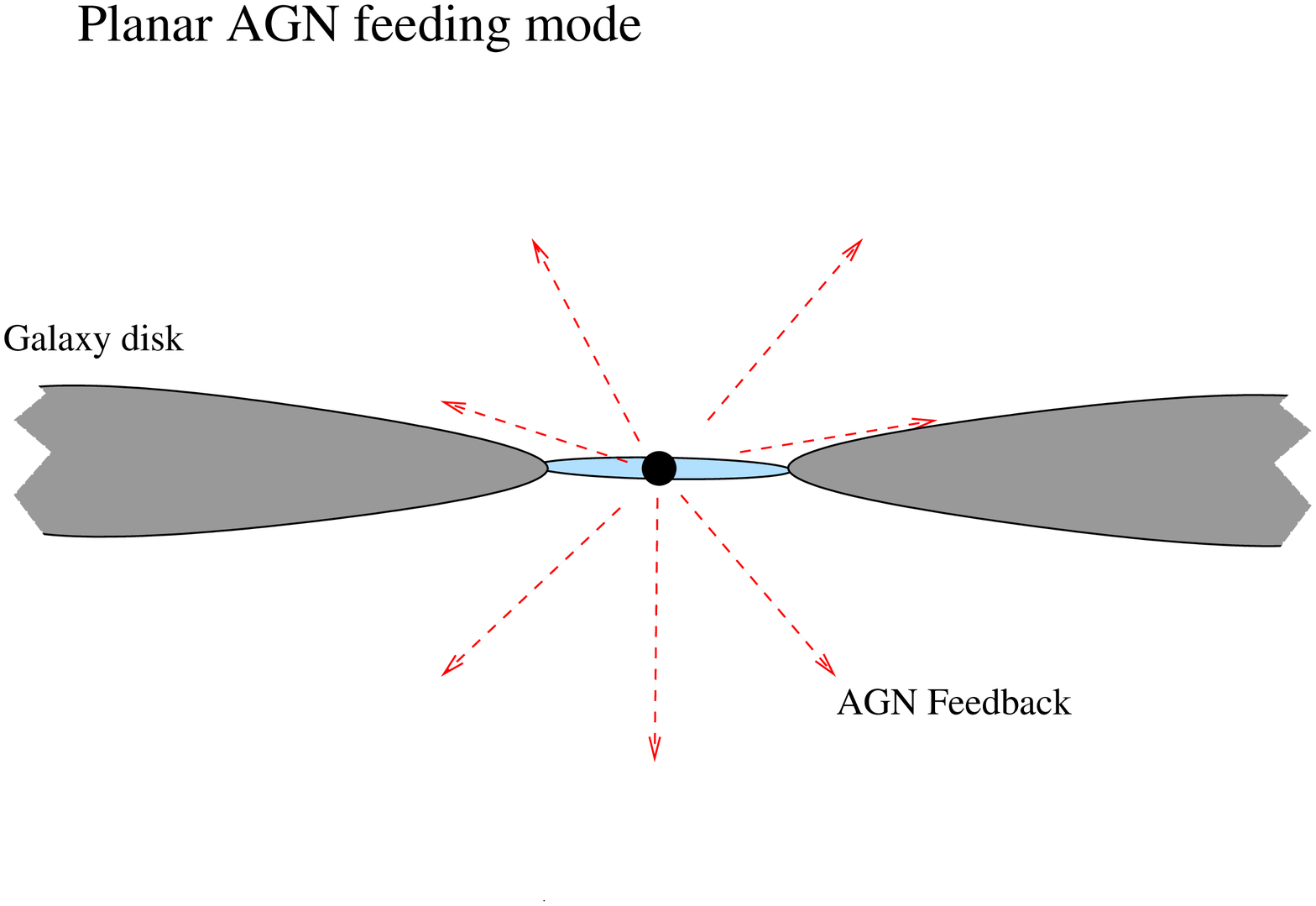,width=.5\textwidth,angle=0}\,\,\,\,\psfig{file=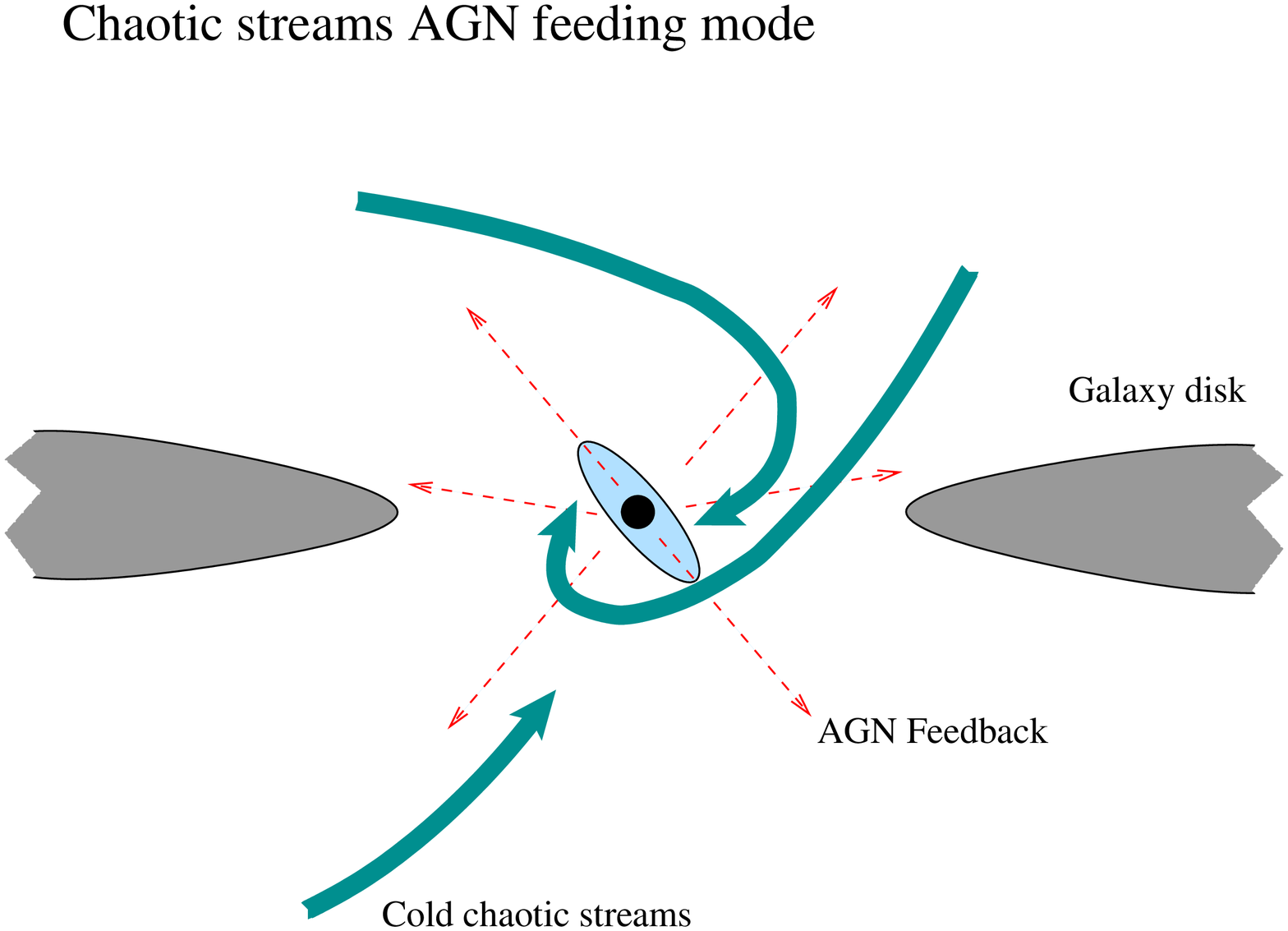,width=.5\textwidth,angle=0}}
\caption{Schematic illustration of the two modes of AGN feeding. Left --
  ``Planar mode'', where gas is delivered from the Galaxy-wide (many kpc) disc
  by a planar inflow through a disc, for example by spiral density waves or
  ``bars within bars''. The inner pc-scale disc (light blue) is always
  oriented the same way as the galaxy disc. Right -- the ``chaotic mode'',
  where the gas is delivered by cold dense filaments or blobs on nearly radial
  orbits. The small scale AGN disc fluctuates in its spin orientation, so that
  the AGN feedback is quasi-spherical when averaged over long ($\sim $ million
  yrs) time scales. The outer Galaxy disc is disconnected from the innermost
  disc by star formation that turns gas into stars before a significant inflow
  could happen.}
\label{fig:fig1}
\end{figure*}

What is missing in the above analysis is the possibility of star formation in
the galaxy disc \citep[this was also not included in the numerical experiments
  of][]{NayakshinPower10}. Massive cool gas discs become gravitationally
unstable when the Toomre $Q$-parameter approaches unity from above
\citep{Toomre64}:
\begin{equation}
Q = {\kappa_{\Omega} c_s \over \pi G \Sigma} \simlt 1 \;,
\label{q-def}
\end{equation}
\noindent where $\kappa_{\Omega} = $ is the epicyclic
frequency. Noting that $\kappa_{\Omega} = \sqrt{2}\Omega$ for the
singular isothermal sphere potential, $\Omega^2 = GM_{\rm total}/R^3$,
and using the hydrostatic balance condition $H = c_s \Omega^{-1}$, we
find that
\begin{equation}
Q = {\sqrt{2} H M_{\rm total}(R) \over R M_{\rm disc}} \;.
\label{q_est}
\end{equation}
\noindent Note that $M_{\rm disc}$, defined just below equation \ref{pdisc}, is function of $R$ in the above 
equation.

It is widely believed that self-gravitating gas discs self-regulate their star formation rates to maintain 
marginally stable states. As a disc becomes self-gravitating, stars form on a local dynamical time scale 
(i.e., extremely rapidly) and the energy and momentum released by the stars heat the disc up; this increases 
$Q$ and the star formation rate is reduced to an equilibrium rate that maintains quasi-equilibrium. Generally, 
the disc {\bf first} becomes unstable when $Q\approx 1.5$, and therefore we arrive at the constraint 
\begin{equation}
{M_{\rm disc} \over H} \approx {M_{\rm total}(R) \over R}\;.
\label{quasi}
\end{equation}
Using this now in equation \ref{mmax0}, we arrive at the ``disc $M_{\rm bh}$--$\sigma$ relation",
\begin{equation}
M_{\rm bh} = {\kappa f_d \over \pi G^2} \sigma^4\;,
\label{new_msigma}
\end{equation}
where we introduce $f_d < 1$ as the gas mass fraction in the disc to account
for the fact that a realistic galaxy disc contains gas and stars as
well. According to this definition, the fraction $1-f_d$ of the disc mass is
in the recently formed stars or very dense massive molecular clumps on the way
to collapsing and forming stars.

We see that the disk $M_{\rm bh}$--$\sigma$ relation is of exactly
same form as the ``spherical" $M_{\rm bh}$--$\sigma$ relation
(equation \ref{msig}) derived by \cite{King03,King05}. However, we
note that the coefficient in front of the latter is smaller than that
in the former. The difference is difficult to quantify exactly, but
would seem to be at least a factor of a few to ten. Physically, it is
due to (1) the factor $f_g=0.16$ in front of the spherical relation
and $f_d$ for the disk relation. It is unlikely that $f_d$ is
significantly smaller than unity during the early gas-rich epoch of
galaxy formation. Furthermore, in principle a similar in meaning
coefficient -- taking into account that part of the spherical shell is
already converted into stars -- may be introduced in the spherical
$M_{\rm bh}$--$\sigma$ relation, in which case the difference between
the two relations would be exactly equal to $f_g$. (2) When deriving
the disc $M_{\rm bh}-\sigma$ relation, we took the most optimistic
assumptions about the AGN feedback on the disc. As discussed below
equation \ref{pdisc}, realistically, a part of the feedback can be
shadowed by the inner disk (the innermost part of which is very hard
to affect by the feedback, see below). Beaming away from the disc
plane may decrease the momentum feedback on the disc further yet.

Another important point is that although our arguments were presented
for a disc geometry, they are also applicable to a bar-mediated gas
inflows, such as those discussed by \cite{Shlosman90}. These authors
argued that ``bars within bars'' may effectively channel the gas into
the central regions of galaxies to fuel AGN. However, bars present
even less of an effective solid angle than discs for AGN feedback to
work on: barred gas inflows are squashed geometrically not only to the
galaxy midplane, as disc inflows, but also in the azimuthal direction
within the galaxy plane. Therefore the momentum striking a bar would
be even less than that for the disc (equation \ref{pdisc}), and thus
we would expect even more massive SMBHs if they were limited by their
feedback on the gas bars only.

The derivation in this section is applicable to regions well beyond the
black hole's influence radius, $R_h = G\mbh/2\sigma^2$, which is typically of
the order of a few to tens of pc. In the Appendix we show that a gas disc at
$R \ll R_h$ is actually able to withstand the momentum feedback of the
AGN. This is why the sketch in Figure 1 shows small scale discs unaffected by
the feedback.

\section{Why do observed SMBHs not follow the disc $M_{\rm bh}$--$\sigma$
  relation?}\label{sec:no_disc}

\cite{KormendyGebhard01} found that SMBHs do not correlate with circular
velocities of galaxy discs. \cite{Hu08} showed that SMBHs in pseudobulge
galaxies are {\em underweight} by a factor of about ten with respect to their
cousins in classical bulge systems. \cite{Graham08} showed that barred
galaxies tend to have SMBHs that are a factor of $\sim 3$ to ten smaller than
elliptical galaxies at the same velocity dispersion. \cite{KormendyEtal11}
found that observed SMBHs do not correlate well with velocity dispersions of
either galaxy discs or pseudobulges.

In the sample of \cite{KormendyEtal11}, only two SMBHs in pseudobulge systems
appear to lie above the classical $M_{\rm bh}$--$\sigma$ relation for bulges,
with the rest being somewhat below the relation, in a qualitative agreement
with \cite{Hu08}. 

If these observations are confirmed in the future, then this is clearly very
strongly at odds with the theory developed above: because discs are difficult to
affect by AGN feedback simply due to geometrical arguments \citep[see also
  simulations by][]{NayakshinPower10}, a SMBH fed by a disc should be able to
grow more massive by at least a factor of a few (compare equation
\ref{new_msigma} to equation \ref{msig}) than a SMBH embedded in a bulge.
In fact, theoretically one may expect disc SMBHs to be even more massive: as noted
below equation \ref{pdisc}, this is the absolute minimum to which the SMBH can
grow if fed through the disc because in general AGN feedback impacting a given
disc radius can be diluted at smaller radii or be beamed away from the disc
(neither of these effects is included in equation \ref{new_msigma}).

The simplest interpretation of this result is to accept that star formation in
large scale discs is too efficient a process and so significant gas flow
into the sub-pc vicinity of SMBH is suppressed, leading to a quenching 
of SMBH fueling \citep[see the arguments by][]{Goodman03,Sirko03,NayakshinEtal07}. 
Therefore, we conclude that these discs do {\em not} extend to the SMBH sphere of 
influence (between 1 and 10 pc, depending on SMBH mass), as suggested in the 
right panel of Figure 1.

\cite{Thompson05} proposed a star-burst disc model which appears to overcome
the difficulties noted by \cite{Goodman03}. Specifically, they argue that the star formation
rate in the disc is limited by the action of energy and momentum feedback from
massive stars and supernovae, allowing some fuel to trickle down all the way
to the nucleus. However, one problem with \cite{Thompson05} argument applied
to the inner galaxy is a mismatch of time scales.  Gravitational collapse of
the disc is expected to take place on the dynamical time scale
\citep{Gammie01},
\begin{equation}
t_{\rm dyn} = \frac{R^{3/2}}{G^{1/2}\mbh^{1/2}} \sim 10^{3} R_{\rm pc}^{3/2}
 M_8^{-1/2} \hbox{years}\;,
\label{tdyn}
\end{equation}
where $M_8$ is the SMBH mass in units of $10^8 \msun$, and $R_{\rm pc}$ is
radius in units of parsec.  \cite{Thompson05} show that the disc cooling
time is much shorter than the dynamical time in their model. On the other
hand, massive stars dominating the feedback release energy on a time scale of
a few million years. Therefore, for radii where $t_{\rm dyn}$ is much shorter
than the lifetime of the massive stars, the feedback might be released too
slowly, i.e., only when the disc would already have collapsed gravitationally.

In conclusion, we believe that the observations of \cite{Hu08,Graham08,KormendyEtal11} imply that
SMBHs do {\em not} grow by a large scale (e.g. tens of pc to $\simgt $kpc) discs.

\section{Chaotic/ballistic feeding of SMBHs}

Surveys show little evidence for correlation between the orientation of jets from AGN and the
large--scale structure of their host galaxies \citep[e.g]{SchmittEtal02,VerdoesKleijn2005}. This shows
that the inner $\simlt$ pc scale gas discs feeding AGN are also uncorrelated
with the large scale discs of host galaxies. In our own Galaxy, the observed
young stellar discs in the Galactic Centre \citep{PaumardEtal06} are inclined
at very large angles to the Galactic plane. This indicates that the disc stars
originated from a deposition of one or two gas clouds with angular momentum
directions very different from that of the Galactic disc
\citep{BonnellRice08,HobbsNayakshin09}. 

An attractive physical picture for AGN feeding that explains these
observations and the results of our paper is that SMBHs in general are
fuelled by stochastic deposition of gas clouds with randomly oriented
angular momentum, as sketched in the right panel of Figure 1.

Numerical simulations by \cite{HobbsEtal11} demonstrated that this situation
may be realised in an initially coherently rotating gas sphere if strong
turbulence (driven by supernova explosions due to star formation in the bulge, for
example) is present. Convergent turbulent gas flows lead to formation of high
density regions travelling through the rest of the gas nearly ballistically. A
small fraction of such regions will be on nearly radial orbits that impact the
innermost parsecs of SMBH directly without going through a large scale disc.
\cite{NK07} argued that such flows create small (pc-scale) warped discs with a
fluctuating orientation. These discs are able to feed the SMBH at much higher
rates than the flat discs \citep[see][]{Goodman03} without becoming
self-gravitating, effectively due to a much faster transfer of gas through the
inner parsecs.

We therefore suggest that the absence of an observed $M_{\rm bh}$--$\sigma$
relation for pure disc galaxies or pseudobulges provides observational support
for the chaotic/ballistic mode of SMBH feeding. In our interpretation of
  the data, the lack of classical bulges in these galaxies shut down the most
  efficient channel by which SMBH grow -- the cold chaotic accretion -- and
  therefore led to their SMBHs growing mainly by planar (disk or bar)
  accretion which we argue is inefficient. Having said this, we note that it
  remains to be seen if $\mbh-\sigma$ relation for chaotic streams conforms to
  the relations obtained by \cite{King03}. 

\section{Deficiencies of our work}\label{sec:def}

We shall now remark on limitations of our work. First of all, we
  only considered the momentum feedback here, whereas more complete treatments
  show that the SMBH outflow switches from momentum driving to energy driving
  at larger radii \citep{King05,CiottiEtal10a}. However, this is not likely to
  affect our conclusions significantly because it is the momentum-driven part
  of the AGN feedback process that constitutes the bottle neck for removing
  the gas to infinity \citep{KZP11}. Furthermore, numerical simulations of
  Zubovas and Nayakshin (2012) that include both the momentum and energy
  feedback forms confirm the points made here on the difficulty of expelling a
  massive dense gas disc for the particular case of the Milky Way.

In common with other analytical papers on the subject
\citep[e.g.,][]{SilkRees98,Fabian99,King03,King05,McLaughlinEtal06} we have
also used an isothermal potential. However the discussion in the end of
section \ref{sec:spherical} shows that this has little effect on momentum
driven outflows studied here. Zubovas \& King (2012, in preparation) have
extended this result to large scale energy-driven outflows.

A more serious worry is that our theoretical model predicts $\mbh \propto
\sigma^4$, whereas different authors now find that $\sigma$ ranges from
somewhat below $4$ to as high as 6 (cf. references in the Introduction). One
possible explanation for this within our momentum-driven feedback model is
that the steeper slope for samples that {\em combine different types of
  galaxies} actually results from a superposition of several $\mbh \propto
\sigma^4$ relations for different galaxies vertically offset in mass Zubovas
\& King (2012, in preparation). Future theoretical and observational work is
needed to resolve this issue.

Finally, given our SMBH feeding focus, we do not attempt to
   relate the gas discs we consider here to exponential {\em stellar} discs
   observed in real galaxies. This would require an additional multi-parameter
   model for transformation of gas into stars and is beyond the scope of our
   paper.

\section{Discussion}
\label{sec:discussion}

Previous analytical treatments of the $M_{\rm bh}$--$\sigma$ relation have
modelled the impact of momentum driven AGN feedback on a spherical gas
distribution \citep[e.g.][]{King03,King05}. We have shown analytically that if
SMBH is fed by a large scale gas disc, there exists a ``disc $M_{\rm bh}$ --
$\sigma$ relation''. The latter has the same shape as the ``spherical''
$M_{\rm bh}$ -- $\sigma$ relation \citet{King03,King05} but is offset to
higher masses by a factor of a few to ten, realistically. 

 We emphasise that the precise value of the offset is model dependent, but the
 general trend is undeniable even for AGN feedback models not considered here,
 such as energy-driven models \citep[e.g.,][]{SilkRees98}. This follows simply
 from the geometrical argument first emphasised by \cite{NayakshinPower10}: a
 dense flat disc and even more so a bar presents a much smaller target for AGN
 feedback to act upon than a quasi-spherical gas distribution, and therefore
 much more massive SMBHs could be built if their masses were limited by a
 feedback argument. We may further allude here to the analogy from the field
 of massive star formation, where radiation pressure was once thought to limit
 stellar growth; instead, recent simulations by \cite{KrumholzEtal09} show
 that accretion through a disc (which we note is stable to self-gravitational
 instabilities on the scales of the simulations) is very efficient as
 radiation is channelled away from the disc. If a similar situation applied in
 galaxy formation, SMBHs would grow as long as there is fuel in the massive
 galaxy-wide discs or bars. One could then expect SMBH becoming comparable in
 mass to galaxy gas discs themselves, which is clearly observationally not the
 case.

Reconciliation of this theoretical result with observations leads to
  interesting and significant conclusions about the dominant mode by which
  SMBHs grow in galaxies. \cite{Hu08} showed that pseudobulge galaxies tend to
  have undermassive SMBHs, and \cite{Graham08} showed that barred galaxies tend
  to have SMBHs that are a factor of $\sim 3$ to ten smaller than elliptical
  galaxies at the same velocity dispersion. While \cite{KormendyEtal11} argue,
  in variance with the above authors, that that SMBHs do not even correlate
  with velocity dispersions of galaxy discs and pseudobulges, we note that
  their results also show that SMBHs in pseudobulge systems are typically
  undermassive with respect to their classical bulge cousins \citep[e.g., see
    Fig. 2 in][]{KormendyEtal11}.

We suggest that these observations hint that SMBHs in most galaxies (1) do not
grow by accretion of gas through $\sim$ Kpc gas discs, most likely because gas
is turned into stars too efficiently, and (2) grow via a mode directly linked
to existence of a classical quasi-spherical bulge. In this case it is natural
that the SMBH in galaxies lacking the classical bulge (pseudobulge galaxies)
are undermassive, and it is also natural that SMBHs do not correlate with
galaxy discs.  For the SMBH accretion modes physically linked to classical
bulges we suggest the stochastic \citep{KingPringle07} or ballistic
\citep{HobbsEtal11} accretion modes in which the bulge is the source of low
and fluctuating in direction angular momentum material.

An additional indirect evidence supporting our conclusions comes from
  recent semi-analytical models of galaxy formation by
  \cite{ShankarEtal12}. These authors show that semi-analytical models which
  turn off SMBH feeding during pseudobulge growth episodes driven by secular
  instabilities (bars) provide a much better fit to the data than models that
  allow a concurrent growth of SMBH and pseudobulges \citep[see Fig. 1
    in][]{ShankarEtal12}. We here thus provide a physical interpretation of
  this empirical result: ``fragmentation catastrophe'' of the outer cold and
  massive regions of AGN discs makes them too susceptible to star formation,
  so that not enough gas trickles down to the SMBH
  \citep{Goodman03,NayakshinEtal07}. 

To avoid confusion, we emphasise that we do {\em not} argue that small (e.g.,
sub-parsec) scale accretion discs are unimportant for SMBH growth.  In the
chaotic accretion picture, there will always be a non-vanishing amount of
angular momentum associated with the accretion flow and this results in the
formation of ``small'' by galactic standards discs \citep[$R\simlt 0.01-0.1$
  pc, depending on the SMBH mass and the accretion rate, cf.][]{Goodman03}
that are resilient to AGN feedback (as we show in the Appendix). Rather, our
key point is that the origin of material that feeds these small discs is
distinct from the larger scale (with respect to the SMBH's sphere of
influence) disc or a galactic bar, and the orientation of these smaller-scale
AGN discs is likely to be chaotic because of the manner in which the
quasi-spherical inflows that feeds these discs fluctuates in time.

We also note that large scale galaxy-wide gas discs and gas bars are obviously
important for galaxy formation as a whole, likely accounting for a significant
fraction of all stars formed and also for a radial redistribution of gas and
stars within the galaxy, but we see no clear evidence that these features are
able to channel their gas all the way to the SMBHs.

\section{Acknowledgments}

Theoretical astrophysics research at the University of Leicester is supported
by a STFC Rolling grant. We thank the anonymous referee whose comments helped
to improve this paper significantly.


\section*{Appendix. Feedback radius.}

We have argued above that AGN feedback can affect a large scale gas disc
expelling the gas outward just as it can for a spherical distribution of
gas. We now show that there exists a critical radius within which this
argument no longer holds: the disc there is resilient to AGN feedback.

Inside the SMBH's sphere of influence, marked by the radius $R_h = G\,M_{\rm
  bh}/\sigma^2$, the gravitational potential is dominated by the SMBH and so
we can replace $M(R)$ with $M_{\rm bh}$. For a SMBH on the $M_{\rm
  bh}$--$\sigma$ relation, the SMBH's radius of influence is given by
\begin{equation}
R_h \approx 10 \mbox{ pc } M_8 \sigma_{200}^{-2}\;.
\label{rh}
\end{equation}
As the result, the critical momentum flux striking the edge of the disc must
be
\begin{equation}
\frac{H}{R} \;\frac{\ledd}{c} \ge {G\,M_{\rm bh}\over R^2} M_{\rm disc}\;.
\end{equation}
Using $M_{\rm disc}\approx \mbh (H/R)$ for a marginally self-gravitating disc again, we arrive at the 
requirement that
\begin{equation}
\kappa \Sigma_{\rm BH} = \kappa {\mbh \over \pi R^2} \le 4\;.
\label{mkappa}
\end{equation}
This can be satisfied at radii greater than 
\begin{equation}
R \ge (\kappa \mbh/4\pi)^{1/2} \approx 25 \mbox{ pc } M_8^{1/2}\;,
\label{r1}
\end{equation}
where $M_8$ is the SMBH mass in units of $10^8 \msun$. Recalling that we assumed that $R\le R_h$, we 
conclude that inside the ``feedback radius'' $R_{\rm fb}$ given by
\begin{equation}
R_{\rm fb} = \min\left[ R_h, (\kappa \mbh/4\pi)^{1/2}\right],
\label{rfb}
\end{equation}
momentum feedback by the SMBH is unable to affect a self-gravitating disc.

Physically the significance of this feedback radius can be appreciated by
considering a major gas mass deposition event -- say, the infall of a very
massive Giant Molecular Cloud or a merger with a gas-rich galaxy that dumps a
significant amount of gas into the inner galaxy. This gas will settle into a
disc because it is very likely to have a non zero net specific angular
momentum. At $R \ll R_{\rm fb}$, the disc cannot be expelled by the AGN,
whereas at $R \ll R_{\rm fb}$ it may as shown in the main body of the paper.

Somewhat counter-intuitively, gas closest to the SMBH is actually hardest to
expel to infinity (although this becomes clearer if one considers that SMBH's
gravity is enormous in the inner $\sim$pc). Deposition of a small cloud at
$R\ll R_{\rm fb}$ may be much more promising for AGN feeding than having a
vast amount of gas at $R\gg R_{\rm fb}$ as the latter is more easily affected by
AGN feedback and star formation.

\label{lastpage}

\end{document}